\documentclass[final,1p,times]{elsarticle} % do not change
\usepackage{graphicx} % do not change
\usepackage{amssymb} % do not change
\usepackage{amsthm} % do not change
\usepackage{lineno} % do not change

\journal{Nuclear Physics A} % do not change
\begin{document} % do not change

\begin{frontmatter} % do not change

%% QM09Author: please enter your  
%% Title, author and address info here; please do not use footnotes

% Your Title - please modify
\title{Recent HBT results in Au+Au and $p$+$p$ collisions from PHENIX}% at RHIC}

% Principle author, and co-authors - please modify
\author{Andrew Glenn$^{a}$ for the PHENIX collaboration}

% Address - please modify
% note that if you have authors from several institutions, we recommend
% labelling these [a], [b], [c] etc.
\address[a]{Lawrence Livermore National Laboratory, % label [a]
7000 East Ave.,
Livermore, CA 94551, USA}

\begin{abstract} % do not change
%% Text of abstract goes here - please modify
We present Hanbury-Brown Twiss measurements from the PHENIX experiment at RHIC for final results for charged kaon pairs from  $\sqrt{s_{NN}}$ = 200~GeV Au+Au collisions and preliminary results for charged pion pairs from $\sqrt{s}$ = 200~GeV $p$+$p$ collisions. We find that for kaon pairs from Au+Au, each traditional 3D Gaussian radius shows approximately the same linear increase as a function of  $N^{1/3}_{\rm part}$.  An imaging analysis reveals a significant non-Gaussian tail for $r \gtrsim 10$~fm. The presence of a tail for kaon pairs demonstrates that similar non-Gaussian tails observed in earlier pion measurements cannot be fully explained by decays of long-lived resonances. The preliminary analysis of pions from $\sqrt{s}$ = 200~GeV $p$+$p$ minimum biased collisions show correlations which are well suited to traditional 3D HBT radii extraction via the Bowler-Sinyukov method, and we present $R_{\rm out}$, $R_{\rm side}$, and $R_{\rm long}$ as a function of mean transverse pair mass.
\end{abstract} % do not change

\end{frontmatter} % do not change

%% QM09: we keep linenumbers at least for initial version
%\linenumbers % do not change

%% start of main text - please modify. Below is a sub-set (single section) 
%% of an earlier proceedings that show how one can handle references 
%% and figures etc.
%%\section{}\label{}

\section{Introduction}
Measurements of the Hanbury-Brown Twiss effect for hadrons in relativistic heavy-ion collisions provide information on the source size and hadron emission duration, allowing a mapping of the freeze-out hyper-surface. Advancement of both HBT experimental analysis methods and hydrodynamic modeling must continue in order to extract precision information, and recent progress in resolving the historic discrepancies between HBT measurements and hydrodynamic model results \cite{Pratt:2008qv} emphasizes this point. The observations of extended, non-Gaussian, source size from two-pion correlations \cite{Adler:2006as} make the measurement of two-kaon correlations important for understanding the contribution from decays of long-lived resonances. 

HBT measurements show azimuthal sensitivity relative to the reaction plane of non-central collisions. Future femtoscopic measurements may similarly use a jet axis to define the event-by-event geometry to gain insight into the modification of jets by the Quark-Gluon Plasma, and conversely, the impact of the jet on the medium. Such measurements will require a better understanding of HBT in $p$+$p$ collisions.

The main source of information for HBT analyses is the two particle correlation function $C_{2}( {\bf q} )$ where, for identical bosons, Bose-Einstein enhancement is manifest as an increased yield at low relative pair momentum {\bf q}. Several forms for $C_{2}( {\bf q} )$ are shown in Eq \ref{eq:eq1}. The first form is 
\begin{equation}
\label{eq:eq1}
A({\bf q} )/B({\bf q} ) = C_{2}( {\bf q} )  \approx  1 + \int d {\bf r} (|\Phi_{\bf q} ({\bf r})|^2-1)S( {\bf r} ) \approx 1 - \lambda + \lambda F_C({\bf q} )[1+G] 
\end{equation}
the experimental correlation function where $A({\bf q} )$ is the measured two particle pair momentum difference distribution and $B({\bf q} )$ is the normalized background distribution. In the second form, the Koonin-Pratt equation,  {\bf r} is pair separation, $\Phi_{\bf q} ({\bf r})$ is the relative wave function and $S({\bf r})$ is the emission source. In the final form, Bowler-Sinyukov, $F_C({\bf q} )$ is the Coulomb correlation and $G = \exp(-R_{\rm side}^2 q^2_{\rm side} -R_{\rm out}^2 q^2_{\rm out}-R_{\rm long}^2 q^2_{\rm long}) $. We discuss this in more detail below.

\section{Methodology}
Charged kaons and pions are reconstructed using the PHENIX Drift Chambers (DC),  Pad Chambers (PC), and lead scintillator Electromagnetic Calorimeters (EMCal). The DC and PC track particles through a magnetic field to obtain momentum, and the EMCal provides time of flight relative to the Beam-Beam Counters (BBC). Matching cuts are applied between detector hits and track projections at the EMCal and DC farthest from the vertex. This information is used to calculate a mass-squared for particle identification. The combination of BBC and Zero-Degree Calorimeters provide vertex and centrality information. 

The experimental correlation function in Eq.~\ref{eq:eq1} is obtained from the ratio of the relative momentum distribution of all pairs within same events, $A({\bf q} )$,  to the normalized relative momentum distribution of pairs from different events within the same event class,  $B({\bf q} )$.  Event class refers to events with similar centrality, collision vertex {\it et cetera}. Pair selection cuts are used to remove track splitting and merging. For the analysis of kaons from Au+Au, Monte-Carlo detector simulations are used to correct for reconstruction inefficiencies for particle pairs with close paths through the DC and EMCal. These inefficient regions are removed by pair cuts for preliminary analysis of pions from $p$+$p$.

Information about the emission source is obtained from $C_2({\bf q} )$ using two methods. The traditional approach makes assumptions about the emission source to derive a form for $C_2$ with parameters which we fit using a log-likelyhood method to best match the measured $A({\bf q} )$ and $B({\bf q} )$. The Bowler-Sinyukov formulation, which assumes a Gaussian emission source at the core and a long range halo, has become the standard formulation of this approach. Using Bowler-Sinyukov requires the choice a coordinate system and reference frame. We choose the out-side-long coordinate system, where long is the beam axis direction, out is the direction of the pair transverse momentum and side is orthogonal to both.  The Longitudinally Co-moving System (LCMS), where total longitudinal pair momentum is zero,  is chosen as the reference frame.  The second approach is to decompose the measured $C_2-1$ using suitable basis functions, and then directly solving the integral equation for $S({\bf r})$ \cite{Brown:1997ku}. This process, referred to as imaging, requires no strong assumptions about the form of $S({\bf r})$, and no explicit Coulomb correction is needed since the full wave function is used.

\section{Charged kaons from Au+Au collisions}
\begin{figure}[bt]
\centering
\includegraphics[height=2.6in]{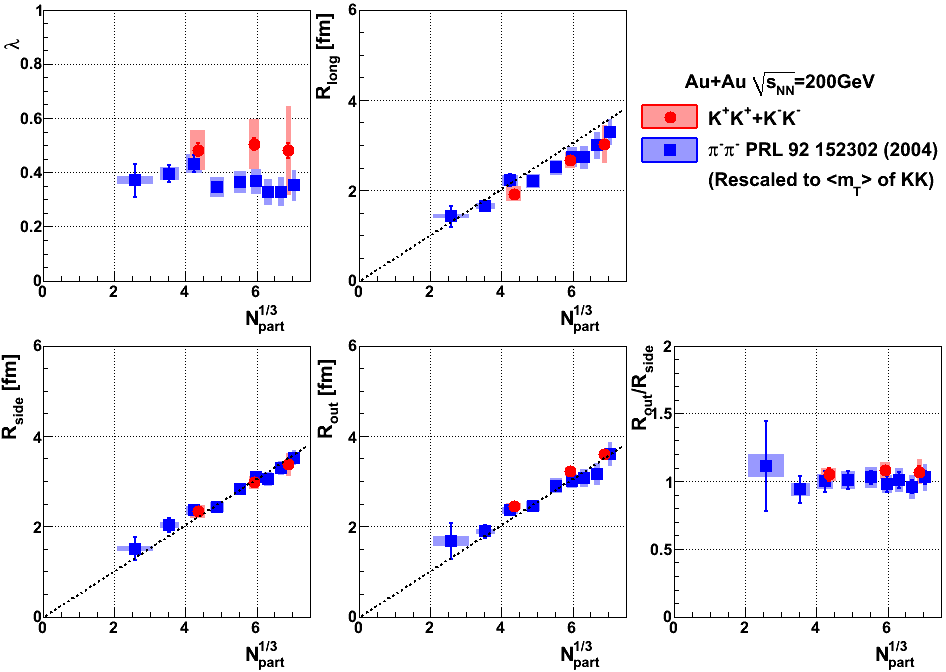}
\caption[]{Centrality dependence of Bowler-Sinyukov source parameters for charged kaon pairs compared to pion data \cite{Adler:2004rq} rescaled to the same transverse mass. The dashed line is the result of a simultaneous fit of all kaon radii to $y=mx$.}
\label{fig:fig1}
\end{figure}

We refer the reader to \cite{Afanasiev:2009ii} for additional analysis details and a full presentation of the data; however, the centrality dependence of the Gaussian source parameters is shown in Fig.~\ref{fig:fig1}. 
Also shown are the pion data \cite{Adler:2004rq} from the same system rescaled to the transverse mass of the kaon data. The rescaled pion data overlap with kaon data, and both show a linear trend. A simultaneous fit of all kaon radii to a line with zero intercept produces a slope of $0.51 \pm 0.01$ and a $\chi^2/$DoF of $14.6/8$. The $m_T$ dependence of the Gaussian source parameters shows poor agreement with traditional hydrodynamic models which match flow and spectra measurements \cite{Afanasiev:2009ii}. A recent 1D hydrodynamic calculation \cite{Pratt:2008qv}, which improves on several aspects of traditional hydrodynamic models and includes novel universal initial flow, reproduces the kaon radii well even though it was not tuned for this data set. The model must now be extended to higher dimensions to test if it can simultaneously describe the flow data. A 1D imaging analysis was also performed \cite{Adler:2004rq}. The reconstructed source shows Gaussian emission at low $r$, but a significant non-Gaussian tail is observed for $r \gtrsim 10$~fm.

\section{Charged pions from $p$+$p$ collisions}
\begin{figure}[bt]
\centering
\includegraphics[height = 1.7in]{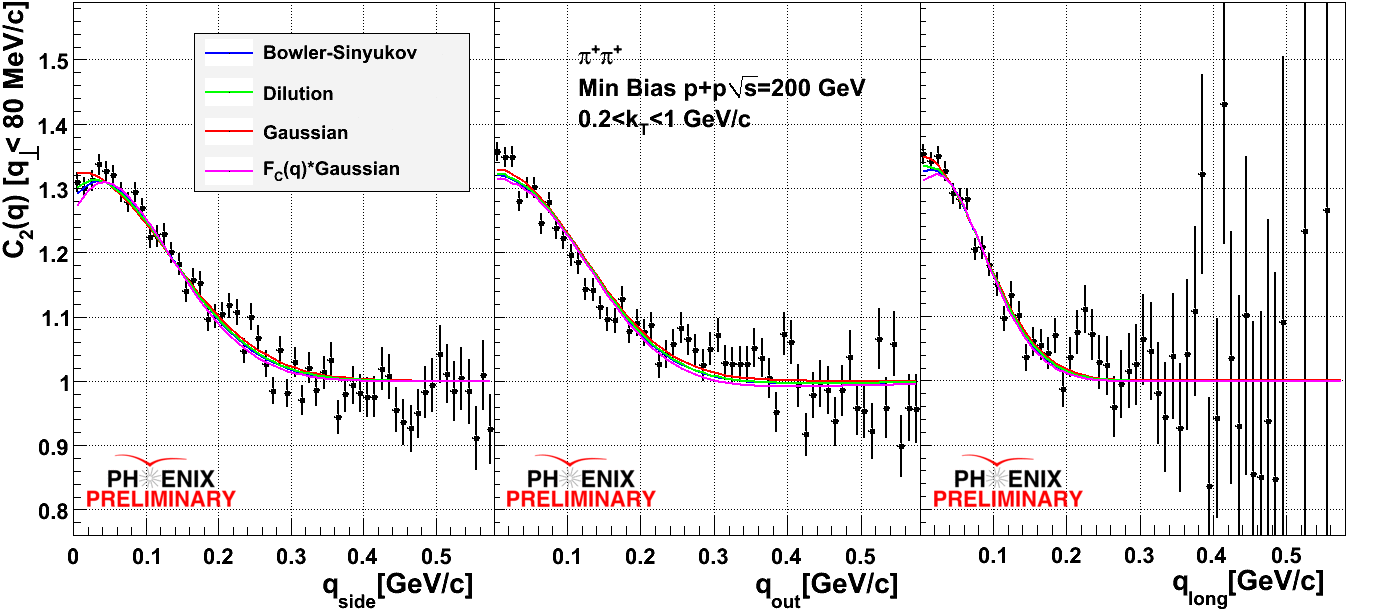}
\caption[]{Slices of the 3D correlation function for $\pi^+$ pairs. Solid curves show results for several fit forms of $C_2$.}
\label{fig:fig2}
\end{figure}

\begin{figure}[bt]
\centering
\includegraphics[height=2.6in]{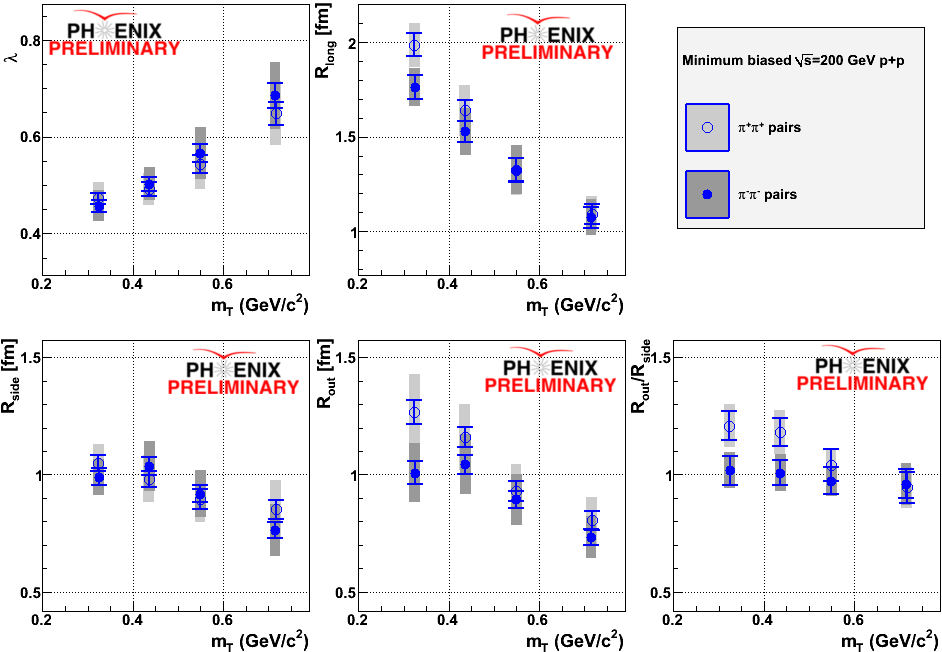}
\caption[]{The transverse mass dependence of the Bowler-Sinyukov source parameters for pion pairs from $\sqrt{s}=200$~GeV minimum biased collisions. We note a correction from the oral presentation, which swapped $R_{\rm out}$ and $R_{\rm side}$ labels.
}
\label{fig:fig3}
\end{figure}

Roughly 2.5 million like sign pion pairs from minimum biased $p$+$p$ collisions from the 2004-2005 RHIC run are analyzed. Despite concerns regarding the smoothness approximation \cite{Pratt:1997pw} and energy-momentum correlations for small and low multiplicity systems \cite{Chajecki:2008vg}, many 1D and 2D correlation slices from $p$+$p$ collisions were examined and verified to be well behaved and sufficiently described by the Bowler-Sinyukov form for our statistics and acceptance. Example 1D slices of the full 3D correlation are shown in Fig.~\ref{fig:fig2}.  
The Bowler-Sinyukov method is used to extract the emission source parameters show in Fig.~\ref{fig:fig3}.
We note a correction from the oral presentation, which swapped $R_{\rm out}$ and $R_{\rm side}$ labels.

\section{Discussion and Outlook}

We find that for kaon pairs from Au+Au, $R_{\rm out}$, $R_{\rm side}$, and $R_{\rm long}$ show approximately the same linear dependence of  $R \approx (0.51\pm0.001)N^{1/3}_{\rm part}$.  An imaging analysis shows that the bulk emission at low radius is consistent with a Gaussian source but a significant non-Gaussian tail is revealed for $r \gtrsim 10$~fm. The presence of this tail for kaon pairs indicates that similar non-Gaussian tails observed in earlier pion measurements have a component other than decays of long-lived resonances, greatly increasing the likelihood of a truly extended emission source.

The preliminary analysis of pions from $\sqrt{s}$ = 200~GeV $p$+$p$ minimum biased collisions show correlations which are surprisingly well suited to traditional 3D HBT radii extraction via the Bowler-Sinyukov method for our statistics and acceptance. An imaging analysis to provide further insight will be performed in the near future.

The next era of HBT should be an exciting time of learning from hydrodynamic models that simultaneously  describe spectra, flow, and HBT as well as experimentally exploring correlations relative to the jet axis in $p$+$p$ and, eventually, heavy ion collisions.

\section*{Acknowledgments} % please check/modify, comment out or delete if not needed
Prepared by LLNL under Contract DE-AC52-07NA27344.

%% end of main text

 % do not change 

\begin{thebibliography}{99} % do not change 
   
  %\cite{Pratt:2008qv}
\bibitem{Pratt:2008qv}
  S.~Pratt,
  %``Resolving the HBT Puzzle in Relativistic Heavy Ion Collision,''
  Phys.\ Rev.\ Lett.\  {\bf 102}, 232301 (2009)
  [arXiv:0811.3363 [nucl-th]].
  %%CITATION = PRLTA,102,232301;%%
 
 %\cite{Adler:2006as}
\bibitem{Adler:2006as}
  S.~S.~Adler {\it et al.}  [PHENIX Collaboration],
  %``Evidence for a long-range component in the pion emission source in Au +  Au
  %collisions at s(NN)**(1/2) = 200-GeV,''
  Phys.\ Rev.\ Lett.\  {\bf 98}, 132301 (2007)
  [arXiv:nucl-ex/0605032].
  %%CITATION = PRLTA,98,132301;%%

%\cite{Brown:1997ku}
\bibitem{Brown:1997ku}
  D.~A.~Brown and P.~Danielewicz,
  %``Imaging of sources in heavy-ion reactions,''
  Phys.\ Lett.\  B {\bf 398}, 252 (1997)
  [arXiv:nucl-th/9701010].
  %%CITATION = PHLTA,B398,252;%%

   %\cite{Afanasiev:2009ii}
\bibitem{Afanasiev:2009ii}
  S.~Afanasiev {\it et al.}  [PHENIX Collaboration],
  %``Kaon interferometric probes of space-time evolution in Au+Au collisions at
  %sqrt(s_NN) = 200 GeV,''
  submitted to {\it Phys. Rev. Lett.}
  arXiv:0903.4863 [nucl-ex].
  %%CITATION = ARXIV:0903.4863;%%

%\cite{Adler:2004rq}
\bibitem{Adler:2004rq}
  S.~S.~Adler {\it et al.}  [PHENIX Collaboration],
  %``Bose-Einstein correlations of charged pion pairs in Au + Au collisions  at
  %s(NN)**(1/2) = 200-GeV,''
  Phys.\ Rev.\ Lett.\  {\bf 93}, 152302 (2004)
  [arXiv:nucl-ex/0401003].
  %%CITATION = PRLTA,93,152302;%%
  
  %\cite{Pratt:1997pw}
\bibitem{Pratt:1997pw}
  S.~Pratt,
  %``Validity of the smoothness assumption for calculating two-boson
  %correlations in high-energy collisions,''
  Phys.\ Rev.\  C {\bf 56}, 1095 (1997).
  %%CITATION = PHRVA,C56,1095;%%

  %\cite{Chajecki:2008vg}
\bibitem{Chajecki:2008vg}
  Z.~Chajecki and M.~Lisa,
  %``Global Conservation Laws and Femtoscopy of Small Systems,''
  Phys.\ Rev.\  C {\bf 78}, 064903 (2008)
  [arXiv:0803.0022 [nucl-th]].
  %%CITATION = PHRVA,C78,064903;%%

  
\end{thebibliography}
\end{document}